**Raman study of the anharmonicity in $YBa_2Cu_3O_x$**

E. Liarokapis
Physics Department, National Technical University, Athens 157 80, Greece

## Abstract

A systematic Raman study in the visible carried out on the $YBa_2Cu_3{}^{16,18}O_x$ (x=6-7) compounds, with isotopic substitution of $^{18}O$ for $^{16}O$, has detected a doping dependent deviation from harmonic behavior for the frequency shift of the in-phase mode, a smaller amount of anharmonicity for the apex mode, and almost no effect for the out-of-phase $B_{1g}$-symmetry phonon. It appears that the amount of anharmonicity depends strongly on the oxygen concentration; it diminishes close to the tetragonal to orthorhombic structural phase transition and close to optimal doping, while it reaches its maximum value for the ortho-II and a tetragonal phase. The almost zero anharmonicity at optimal doping persists even at 77K. The data in the overdoped oxygen concentration, where a softening of the in-phase phonon frequency occurs, indicate that the anharmonicity is not enhanced by the sudden increase in the $CuO_2$ buckling. The results fully agree with recent studies of the ortho-II phase but they do not comply with a static double-well potential of the apical oxygen atom at optimal doping.

## 1. Introduction

It is now thirty years since Georg Bednorz and K. Alex Müller surprised the scientific community with their discovery of a superconductor with a transition temperature ($T_c$) above 30K [1], opening the way for the discovery of other compounds with $T_c$ above the liquid nitrogen temperature [2]. At that time it was generally believed that 30K was strictly an absolute upper limit for $T_c$, and the majority of the scientists had abandoned the search for high temperature superconductors (HTSC). It was the discovery of the new layered superconductor that revived the interest on this field. Right after the discovery it was found that the new family of superconductors (the cuprates) had unusual properties quite different from the conventional superconductors. The existence of magnetism and a Mott transition at low doping [3], the d-wave symmetry of the energy gap [4], the very small isotopic shift of $T_c$ at optimal doping [5], the existence of a pseudogap (PG) [6], etc, have shown that this new class of compounds had exotic properties, quite different from the conventional superconductors. The great majority of the scientists being certain that electron-phonon interaction, that was the basic mechanism for superconductivity in the BCS theory, could not explain such $T_c$ values found in the cuprates, have assumed that other alternative coupling mechanisms must be present, and the lattice should not play a role in the process. But, it was found that the isotopic shift of the transition temperature was strongly dependent on the amount of doping [7]. Besides, isotope effects were discovered on the magnetic penetration depth [8], the pseudogap temperature (T*) [9], etc, showing a substantial involvement of the lattice in the processes and pointing to a polaronic approach to superconductivity.

An explanation of the small oxygen isotope effect of $T_c$ observed in the ceramic superconductors, within the frame of the phonon mediated pair coupling mechanism, was based on the enhancement of the electron-phonon coupling strength through the anharmonic motion of the oxygen atoms [10] in a double-well potential [11]. Among the different oxygen sites of $YBa_2Cu_3O_x$, the apex oxygen ($O_{ap}$) has been observed to move in a double well potential [12], and the relative $A_g$-symmetry mode was detected anharmonic [13]. A detailed investigation of the effect of the oxygen isotopic substitution on the phonon modes of $YBa_2Cu_3O_x$ over the whole range of doping (x=6-7) revealed that, among the modes that involve motion along the c-axis of the oxygen atoms, the $B_{1g}$ symmetry out-of-phase vibrations of the plane oxygen atoms ($O_{pl}$) follow well the harmonic approximation, while the in-phase vibrations of $O_{pl}$ and the apical phonon present doping dependent deviation from the harmonic approximation [14]. Recent investigations in $YBa_2Cu_3O_{6.5}$ have found strong indications for the double-well potential of the apical oxygen in the ortho-II phase [15] in complete agreement with the Raman data. On the other hand, the anharmonicity observed by Raman spectroscopy seems to diminish close to optimal doping (increasing towards the ortho-II phase) and it is stronger for the in-phase vibrations of $A_{1g}$-symmetry of the plane oxygen atoms, than for the apical one [14].

This appears to contradict and question the double-well potential picture observed at optimal oxygen concentration [12].

In this work the results of the Raman measurements on the $YBa_2Cu_3{}^{16,18}O_x$ compounds in the full range of oxygen doping will be reviewed and compared with measurements that strongly indicate the motion of the apical oxygen in a double-well potential. The idea of these measurements was entirely due to Prof. Dr. K.A. Müller interested to obtain information about the amount of anharmonicity in the cuprates in the whole range of oxygen concentration. This required samples of excellent quality that have been provided by the group in ETH of Prof. Dr. E. Kaldis.

## 2. Experiment

Two series of samples of $YBa_2Cu_3O_x$ (x=6-7) have been examined, with the isotopic substitution of $^{16}O$ by $^{18}O$ around 90-94%. The samples were prepared under similar conditions with those described before [13]. The amount of oxygen in the YBCO samples has been measured with a volumetric method [16] with an extremely high accuracy ($\Delta x = \pm 0.001$). No other phases have been detected by x-ray powder diffraction and Raman measurements. Raman spectra were obtained with a T64000 Jobin-Yvon triple spectrometer equipped with a liquid nitrogen cooled charge coupled device (CCD) and a microscope (magnification ×100). The 514.5 and 488.0 nm and the 647.1 nm line of an $Ar^+$ and a $Kr^+$ lasers respectively were used for excitation at a power level of ~0.15 mW on the sample, while the laser spot diameter was of the order of 1-2 μm. Local sample heating due to the laser beam was estimated to be <10K. The spectra were obtained from individual micro-crystallites in the approximate $y(zz)\bar{y}$ (or $x(zz)\bar{x}$) and $y(xx)\bar{y}$ (or $x(yy)\bar{x}$) scattering configurations as the x and y axis could not be discriminated in the twinned samples. Accumulation times were of the order of 1 to 3 hours and several micro-crystallites have been examined for each oxygen concentration. The proper orientation of the microcrystallites was selected from their shapes and from the scattering selection rules.

## 3. Results and discussion

The strong modes of $YBa_2Cu_3O_x$ are those of $4A_g$ and $1B_{1g}$-like symmetry and involve vibrations along the c-axis of the Ba atom, the Cu of the $CuO_2$ planes ($Cu_{pl}$), the apical oxygen, and the two vibrations, in-phase ($A_g$) and out-of-phase ($B_{1g}$) of the plane oxygen atoms. Characteristic spectra for selected concentrations and the two isotopes have been presented elsewhere (Figs.1-2, Ref.[14]). Apart from a frequency shift of the oxygen modes due to the isotopic substitution, the two sets of spectra looked the same. No frequency difference is observed between the two isotope compounds regarding the Ba and the $Cu_{pl}$ phonons due to the negligible amount of coupling between the motion of the oxygen atoms and the vibrations of the Ba, $Cu_{pl}$ atoms [14]. Fig.1a,b,c show the frequencies of the three oxygen modes ($B_{1g}$, in-phase, and apex) for the whole range of oxygen doping for $^{16}O$ and for $^{18}O$, corrected according to the square root of the mass ratio, for two $^{18}O$ contents, i.e., 90% and 94%. The amount of the $^{18}O$ substitution for $^{16}O$ was estimated to be equally distributed to all oxygen sites [17], that will induce a frequency shift of 5.5% (for 90% or 5.7% for 94%) according to the average mass ratio of the two isotopes, or slightly smaller (5.3% for 90%) with lattice dynamic calculations [18]. Theoretical calculations have shown that the oxygen isotope correction factor may slightly depend on the phonon modes [18] or on the oxygen concentration [19], but the changes expected are small. Since the error in the frequency shift from the uncertainty in the exact amount of oxygen substitution (90-94%) is larger than the difference between the simple mass ratio law and the lattice dynamic calculations, we have used the former simple equation in correcting the frequencies for the $^{18}O$ data (Fig.1a,b,c).

One can induce from the data that the out-of-phase $B_{1g}$-symmetry oxygen mode (observed in xx-polarization) for both correction factors (5.5% or 5.7%) agree within statistical error (mean square deviation of frequency values obtained from many microcrystallites), in most of the oxygen doping range, except for a small deviation ($<2cm^{-1}$) close to tetragonal to orthorhombic structural phase transition (x~6.35) and possibly in the overdoped region (Fig.1a). These results agree also with those from IR measurements [13]. The deviation varies with the amount of isotopic mass correction and it could be due to the correction factor that may vary with doping [19]. For both isotopes, there seems to be a small non-linear dependence of the $B_{1g}$ frequency on the oxygen doping level (Fig.1a). Since the frequency of the $B_{1g}$ mode does not vary appreciably with the amount of oxygen, one could use the average values of the two frequencies for $^{16}O$ and $^{18}O$ in order to estimate the average correction factor, which turns out to be very close to the amount of $^{18}O$ substitution measured by other means (≈90%).

For the in-phase mode of the same $O_{pl}$ atoms (of $A_{1g}$-symmetry) the deviation is much stronger (up to 5 $cm^{-1}$) and depends on the amount of oxygen concentration. Again, depending on the chosen isotopic correction, the data from the two isotopes agree quite well close to optimal doping. Even the softening of the in-phase mode, discovered in the overdoped region [20] has the same behavior (sudden drop around optimal doping) for both isotopes (Fig.1b). The maximum amount of anharmonicity seems to appear around the ortho-II phase (x=6.5) and for the tetragonal one (x<6.2). Due to the uncertainty of the exact amount of substitution and the possible dependence of the correction factor on oxygen doping, it is difficult to obtain exact information on the amount of anharmonicity.

Fig.1c presents the data for apical oxygen ($A_{1g}$ symmetry). The difference between the two isotopes is smaller than for the in-phase mode and depends again on oxygen doping and the correction factor. One can generally induce that there is some amount of anharmonicity that disappears in the tetragonal phase (x<6.2). In the optimally doped region, depending on the correction factor, there might be anharmonicity. From the data it becomes obvious that since we are dealing with small anharmonicity, we should find a way to bypass the uncertainty in the evaluation of the amount of $^{18}O$ substitution for $^{16}O$ and the correction factor. Assuming that there is an equal amount of $^{18}O$ substitution in the planes and the chains, we can take the ratio of the frequencies for both isotopes and avoid this uncertainty. For the in-phase mode, the assumption of equal amounts of site substitution is not needed, but it remains a possible uncertainty about the correction factor for the two phonons from the same atoms but of different symmetry.

Fig.2a presents for the two isotopes the ratio of the in-phase and $B_{1g}$ phonon frequencies that involve vibrations of the plane oxygen atoms. Since the frequency ratio does not depend on the exact amount of $^{18}O$ substitution and the correction factor, the statistical variation of the data has been reduced and more accurate conclusions can be drawn. Although the ratio of the $^{18}O$ data varies smoothly with doping, there is a remarkable deviation for the $^{16}O$ compounds (Fig.2a). The average variation of the ratio is presented with the dashed line. It is clear that the deviation from the harmonic potential is almost zero towards optimal doping and increases with the reduction of the amount of oxygen. Besides, there is a strong fluctuation in the amount of anharmonicity, which becomes very small or zero at $x\approx6.3$ and for $x\geq6.85$. Its maximum value occurs for $x\approx6.5$ (apparently at the ortho-II phase) and for $x\approx6.15$ (a tetragonal phase). The anharmonicity seems to disappear close and above the optimal doping. In the overdoped region ($x\geq6.94$), where the in-phase mode was found to soften suddenly [20], the anharmonicity in the compound remains very small or even zero. The similarity of the in-phase mode in the two isotopes can be clearly observed in Fig.3, where the frequency is plotted vs the transition temperature. Since the softening of the phonon was correlated with a change in the $CuO_2$ buckling, it is induced that this deviation from flatness of the $CuO_2$ is not related with an increased anharmonicity.

The corresponding ratio of the apical phonon mode frequency to the $B_{1g}$ is presented in Fig.2b. As discussed before, in this case there is an uncertainty on the relative amount of site selective oxygen substitution, although the controllable methodology followed in sample preparation points to an equal amount of substitution at the plane and apical oxygen sites [17]. A similar trend of anharmonicity is observed as in the case of the in-phase mode of Fig.2a. Close to the ortho-II and the tetragonal phases the amount of anharmonicity appears to become maximum and it reduces towards optimal doping, although in this case there is a small amount of anharmonicity for $x>6.85$. In the case of apex phonon, the data for the two sets of isotopes coincide at $x\approx6.35$, i.e., close to the tetragonal to orthorhombic (T-O) structural phase transition (Fig.2b). Considering the statistical errors, we can safely assume that for both phonons (apex and in-phase) the anharmonicity vanishes at the T-O transition. The small downshift in frequency at the overdoped region appears to be the same for both isotopes, as was the case for the in-phase mode.

The above results disagree with those of Ref. [13] concerning the apex oxygen (in that work the in-phase mode could not be studied). This difference could be attributed to the different excitation wavelength used in Ref. [13]. It has been shown [20] that mainly three phases coexist and contribute to a different amount to the apex mode (Fig.4), and a resonance behavior for the phases of the apex mode has been also established [21]. If we therefore assume, as our data indicate, that the degree of anharmonicity varies considerably among the phases (maximum for ortho-II and tetragonal, minimum for the ortho-I), and take into consideration the resonance of the other phases with increasing laser wavelength excitation [21], the deviation between our data and those of Ref. [13] can be explained. The excitation with IR will enhance the anharmonicity as we approach optimal doping, because of the increased contribution of the anharmonic phases to the Raman signal. It is possible that the small anharmonicity observed in our data in the apex mode (Fig.2b) express small contribution of the other anharmonic phases.

In the ortho-II and other phases there will be a folding of the unit cell and, therefore, more phonons will be Raman active [22]. One could expect the new extra phonons at the Γ-point of the ortho-II phase to either induce new modes at the side of the existing phonons of ortho-I phase or, being very close in frequency, modify their width and intensity. New modes appear mainly for the Ba mode [14], but not close in frequency to the apex phonon, which retains almost the same width with the ortho-I phase (Fig.5). These results agree with the theoretical investigations [22], which predict weak modes not far in frequency for the apical phonon. Since the width of the apical phonon in ortho-II (and the tetragonal) is as narrow as for the ortho-I phase, it is unlikely that the phonon frequency modifications with isotopic substitution to be related with new modes that appear at the Γ-point, although it cannot be excluded.

As described, our data do not seem to support any strong anharmonicity of the apex oxygen close to optimal doping, disagreeing with the conclusions of Ref. [12] that the apical oxygen moves in a double well potential, occupying two distinct sites quite apart (0.12Å distance). Since their measurements had been carried out at low temperatures, we have tested the anharmonicity at near optimal doping 77K. The results are shown in Fig.6 and it is clear that the two isotopes follow the same trend, being independent of temperature. Therefore, even these preliminary data at low temperature show that there is no appreciable change in the amount of anharmonicity for the region 77-300K and, therefore, the different temperature of our measurements with those of Ref.[12] cannot be the reason of the discrepancy. The motion of the apical oxygen in two potential wells does not agree also with the phonon width, which remains smaller in the ortho-I, the ortho-II and the tetragonal phases (Fig.5), increases substantially in the intermediate doping levels where many phases coexist (Fig.4) and decreases, as expected, at low temperatures. Furthermore, the distance of 0.12Å between the two sites of the apical oxygen should correspond to a substantial shift of the apical oxygen phonon. Based on the data of the apical phonon frequency along the c-axis $A_{1g}$-symmetry for the whole range of doping and the measured variation of the apical oxygen position [14] one would expect a shift of ~20 $cm^{-1}$ between the phonon frequencies from the two wells. This is comparable with the total width of the phonon mode at optimal doping (Fig.5), which decreases with temperature. Therefore, the discrepancy between the Raman measurements and those which find a double well potential in the ortho-I phase remains.

For the ortho-II phase it is clear from our data that there is anharmonicity, which appears stronger for the in-phase mode, it is smaller for the apical oxygen atom and almost absent for the $B_{1g}$ phonon. It is likely that the in-phase vibrations of the $CuO_2$ planes are affected more than the out-of-phase ones ($B_{1g}$ mode) from the apical motion that modify the amount of charge transfer. In any case our data agree with other measurements that have found strong anharmonicity for the ortho-II phase [15]. In our data we have found equally strong anharmonicity for a tetragonal phase, which must correspond to a low oxygen concentration and should be checked by other techniques as well. It should be also marked the absence of ahnarmonicity at the tetragonal to orthorhombic structural phase transition (Fig.2). Since the softening of the modes relaxes at this T-O transition, the anharmonicity could be related with parallel to the plane modifications, as other measurements indicate [23].

Concerning the discrepancy between the Raman data and other measurements for the optimally doped samples, one possible explanation can be based on the assumption that Raman active phonons at the Γ-point cannot detect the lattice distortion as the long wavelength phonons will feel only the average position of the oxygen atoms. Then, we have to accept the view that the anharmonicity observed in the ortho-II and the tetragonal phase is due to the contribution of the folded phonons from the edge of the Brillouin zone back to its center or that the effect is dynamic and at low doping level evolves towards a static disorder.

## 6. Summary

In conclusion the presenter Raman data indicate that there is anharmonicity for the in-phase vibrations of the $O_{pl}$ oxygen that depends strongly on the amount of oxygen, being maximum in the ortho-II phase and another one in the tetragonal structure of the compound. At optimal doping the amount of anharmonicity diminishes, and the softening of the mode in the overdoped region seems to scale according with the mass law. For the apical oxygen the problem is more complicated due to existence of many contributing phases, which appear to have a different amount of anharmonicity, with the ortho-I phase being almost harmonic and the ortho-II phase the most anharmonic. The Raman data do not agree with a static double well potential for the apical oxygen at least in the optimally doped region. It is possible that phonons at the Γ-point cannot detect this lattice distortion, which can only be observed from edge phonons of the Brillouin zone.

## 7. Acknowledgments

I would like to thank Prof. Dr. K.A. Müller who had proposed this experimental Raman study, Prof. Dr. E. Kaldis and Dr. K. Conder who provided the excellent quality samples, and Dr. N. Poulakis and Dr. D. Palles who had carried out the time consuming Raman measurements.

**Figure captions**

Fig.1 The variation of the frequency of the $B_{1g}$-like (a), in-phase (b), and apical (c) oxygen phonons with the amount of doping x for the $^{16}O$ and the $^{18}O$. The data for $^{18}O$ have been normalized according to the two isotopes average mass ratio (for 90% and 94% substitution). The continuous lines are guides to eye.

Fig.2 The variation of the relative frequency of the in-phase (a) and the apical (b) to $B_{1g}$-like phonon for all oxygen concentration x and the two isotopes. The continuous lines are guides to eye. The broken line in Fig.2a indicates an average (2$^{nd}$ order polynomial curve) of the data.

Fig.3 The dependence of the in-phase phonon frequency on $T_c$ for the two oxygen isotopes.

Fig.4 The relative amount of the three phases (ortho-I, ortho-II, and tetragonal) for the $^{16}O$ isotope, as induced by deconvolution of the apical phonon. A similar curve is obtained for $^{18}O$.

Fig.5 The doping dependence of the apex and in-phase phonon width for the two isotopes.

Fig.6 The variation of the relative frequency with temperature of the in-phase and the apical to $B_{1g}$-like phonon for optimally doped compound for both oxygen isotopes.

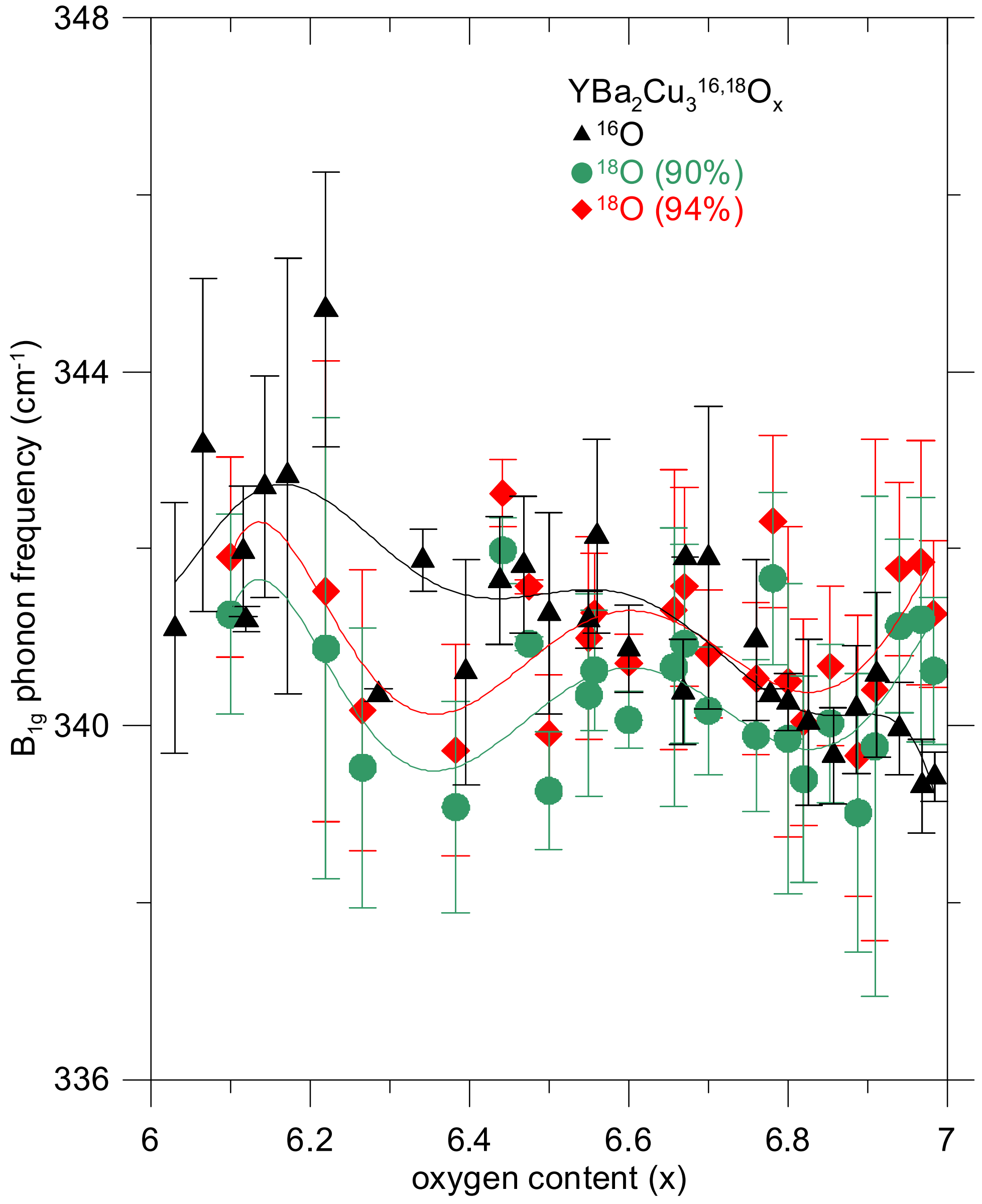

Fig.1a

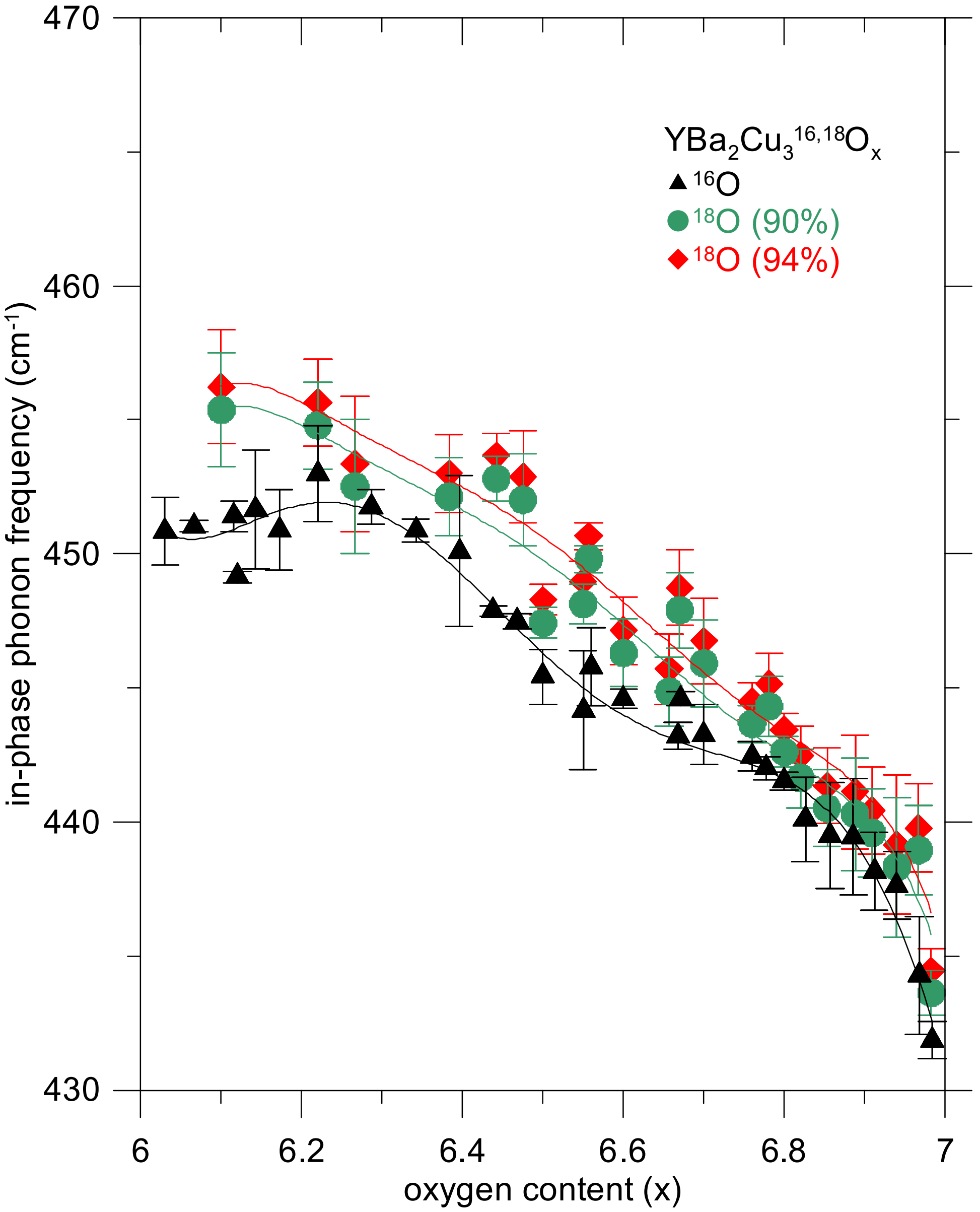

Fig.1b

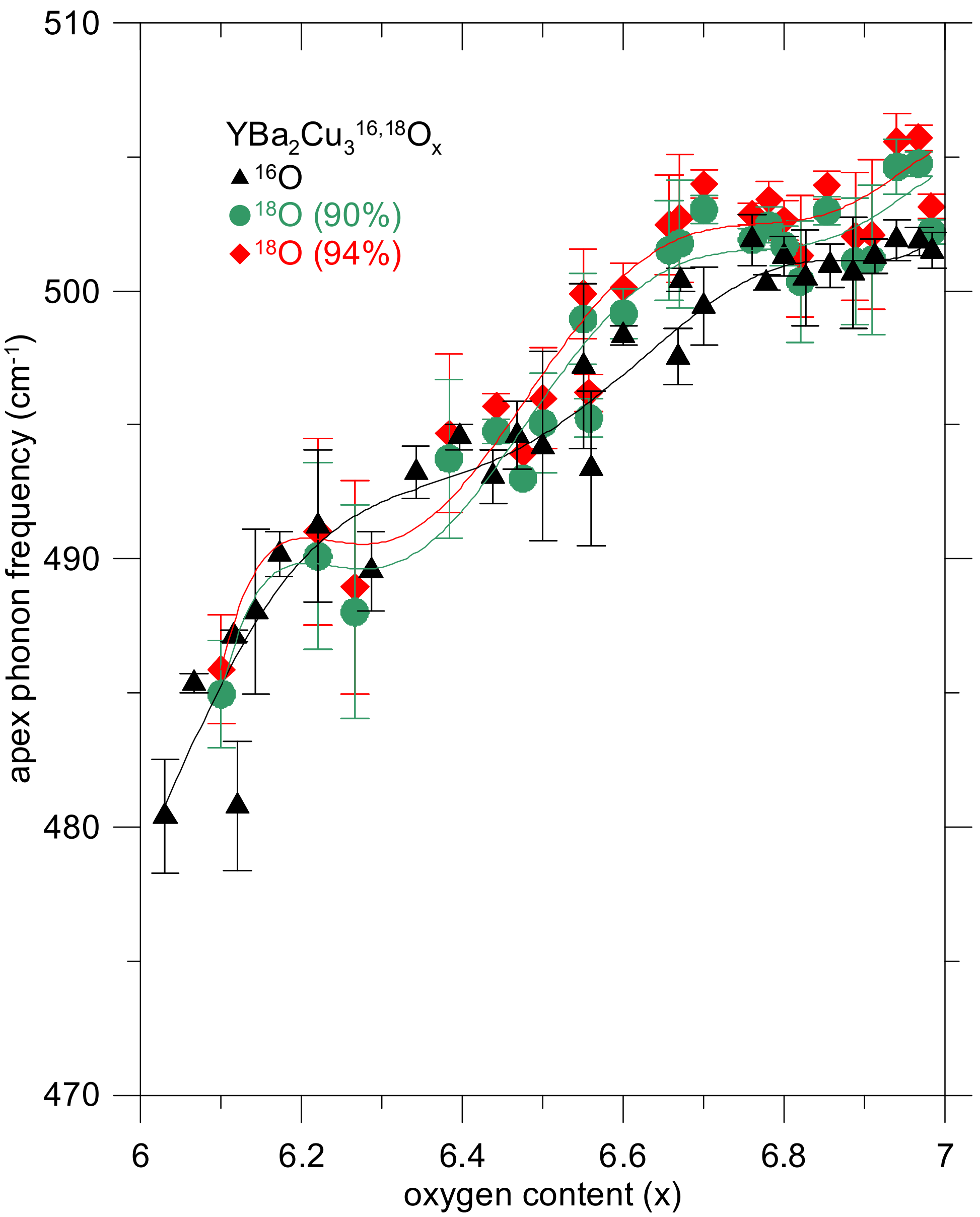

Fig.1c

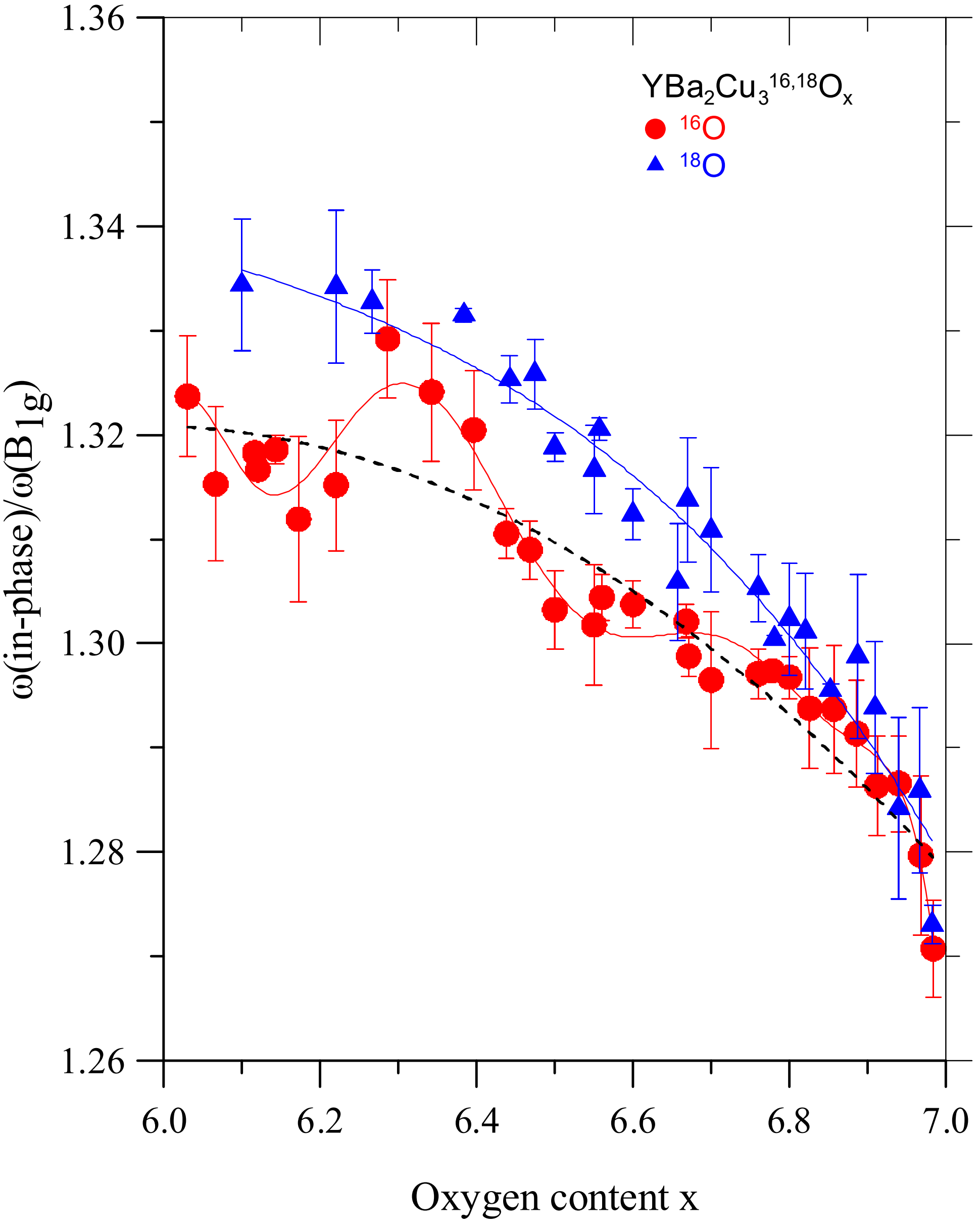

Fig.2a

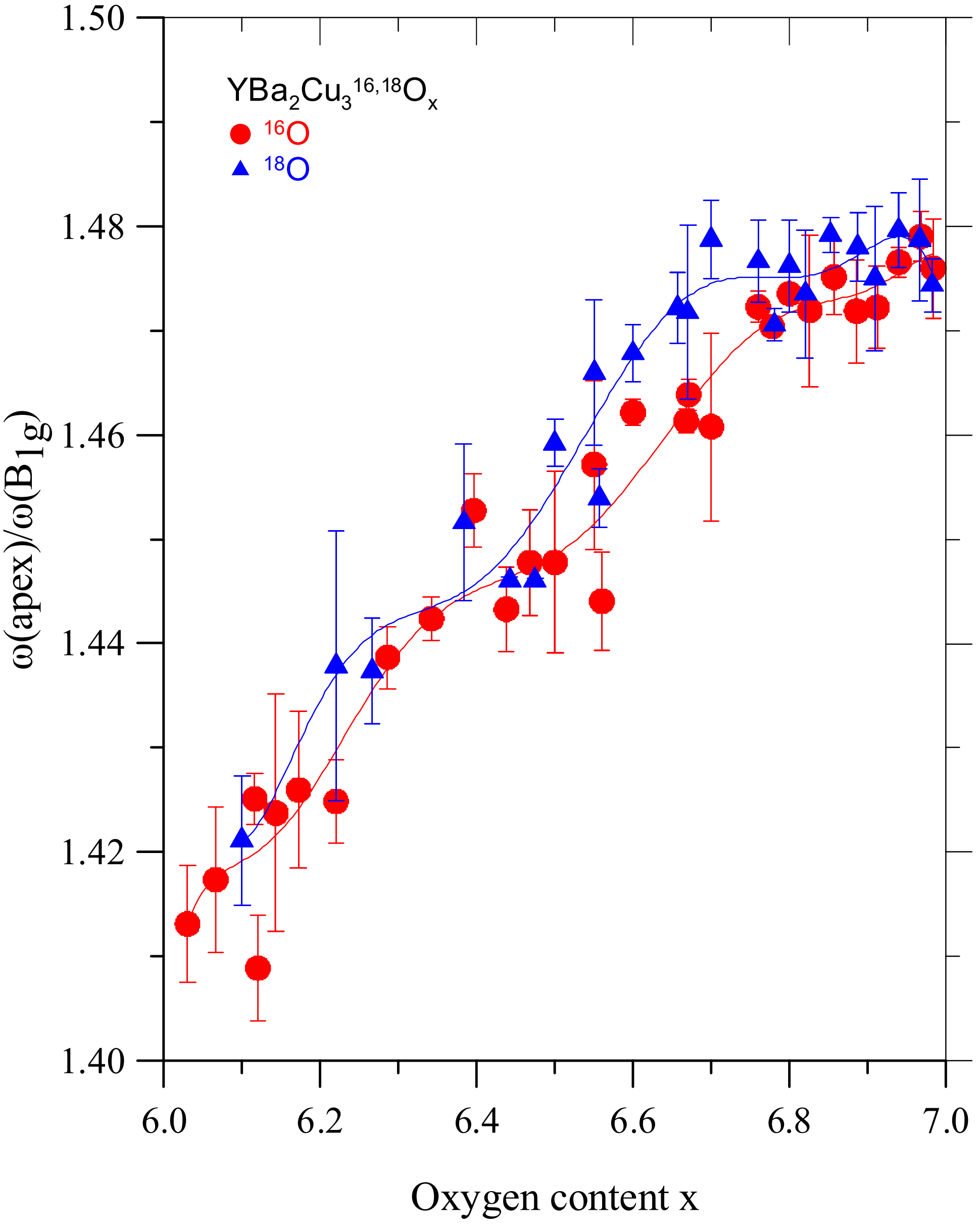

Fig.2b

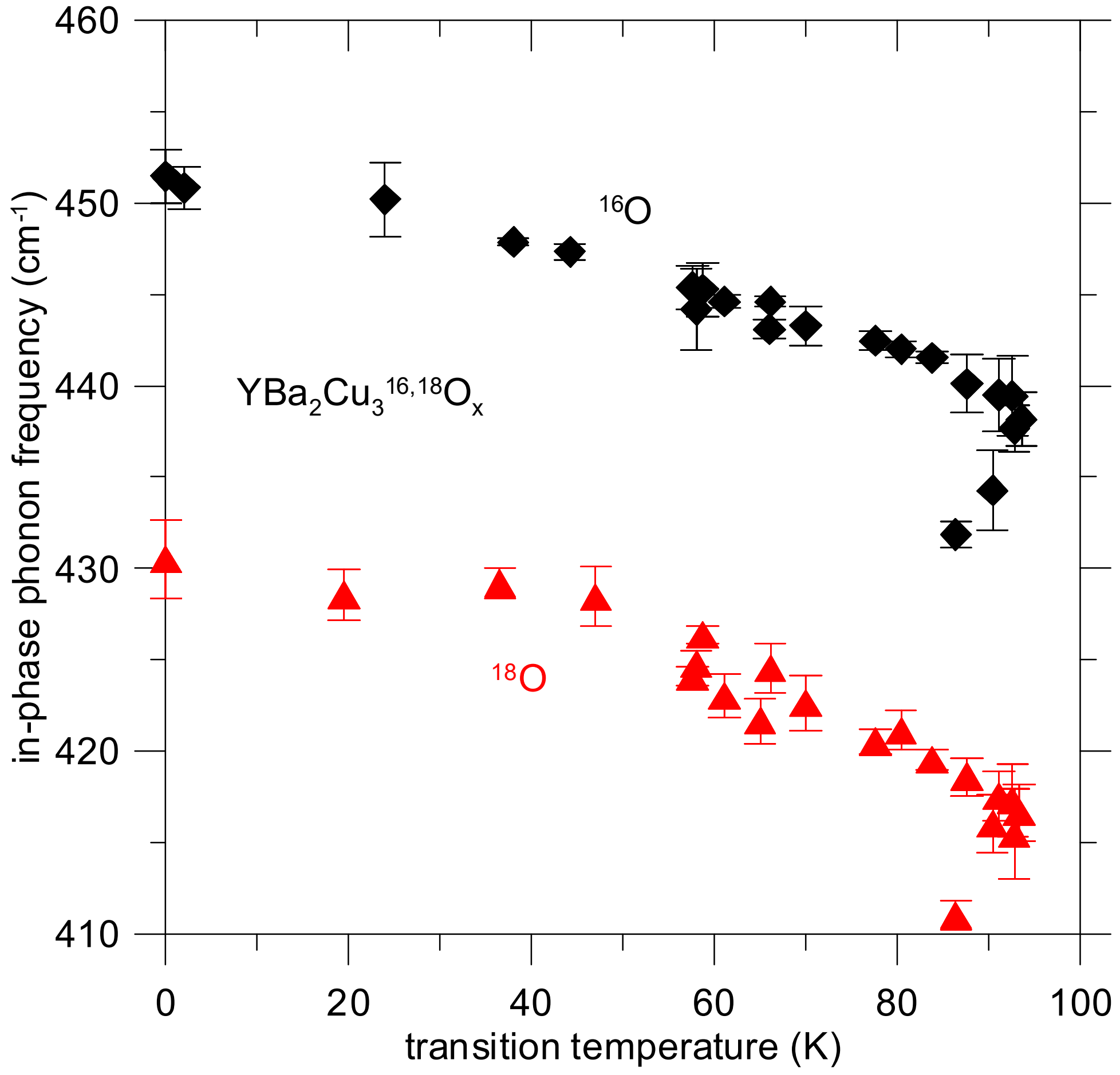

Fig.3

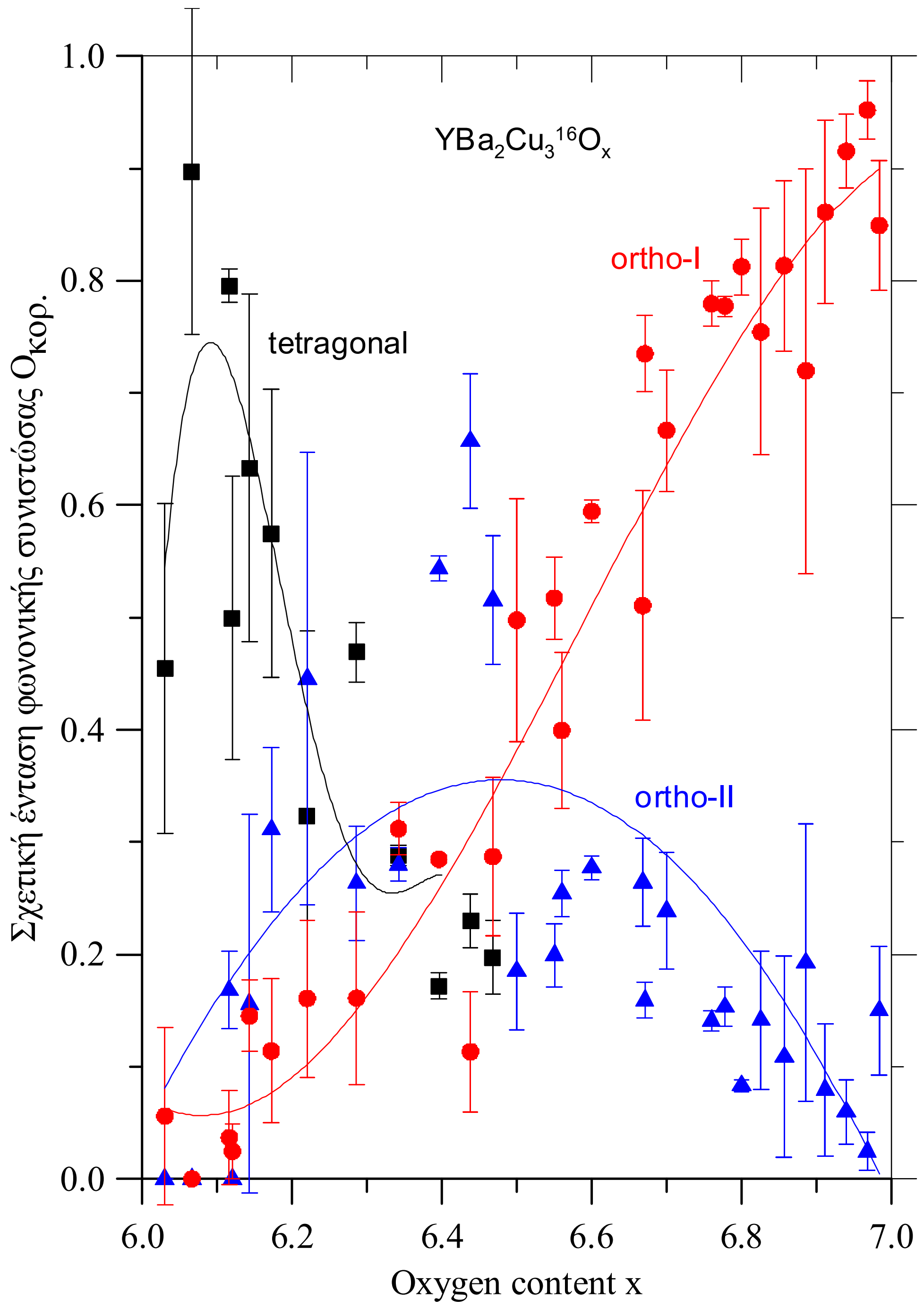

Fig.4

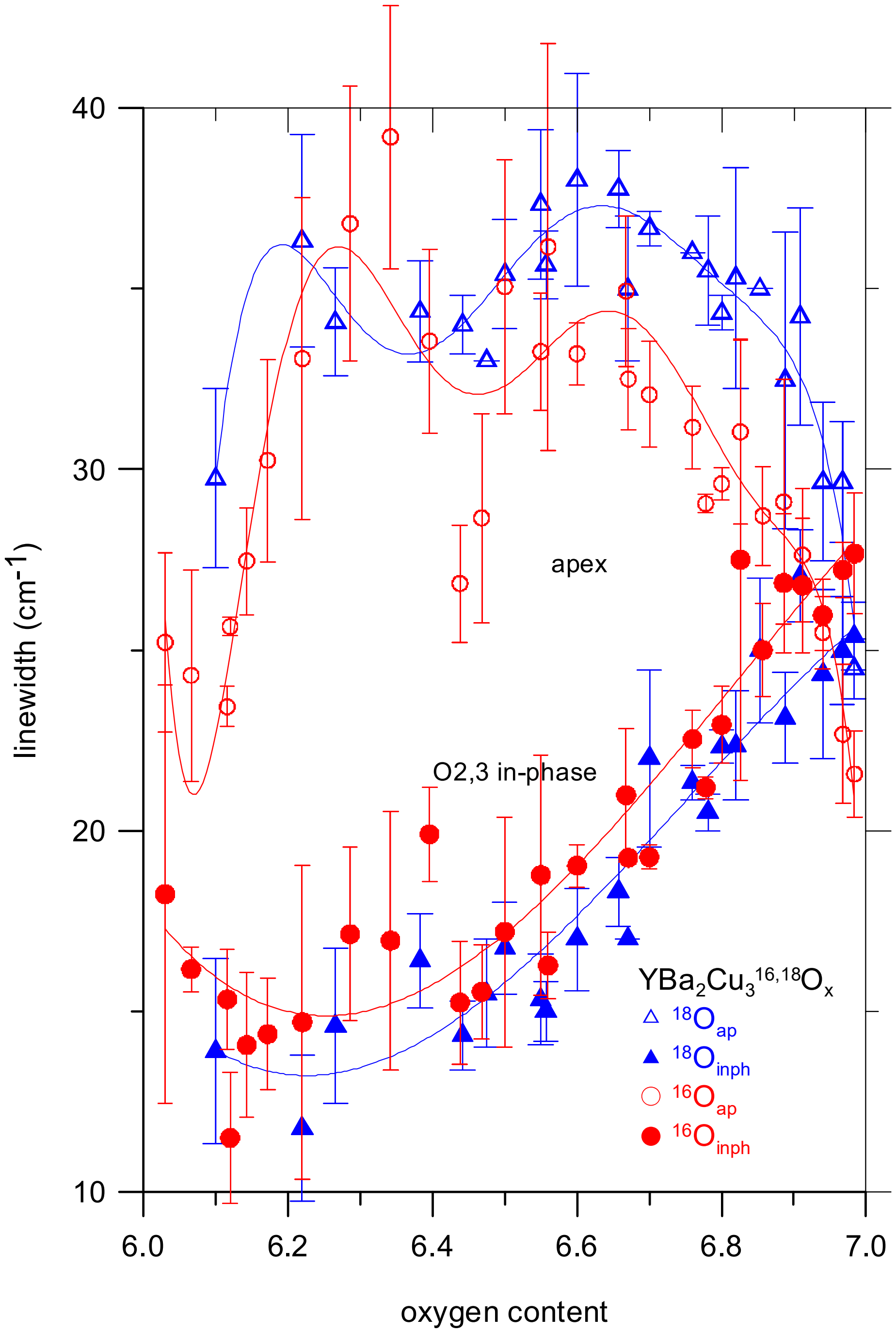

Fig.5

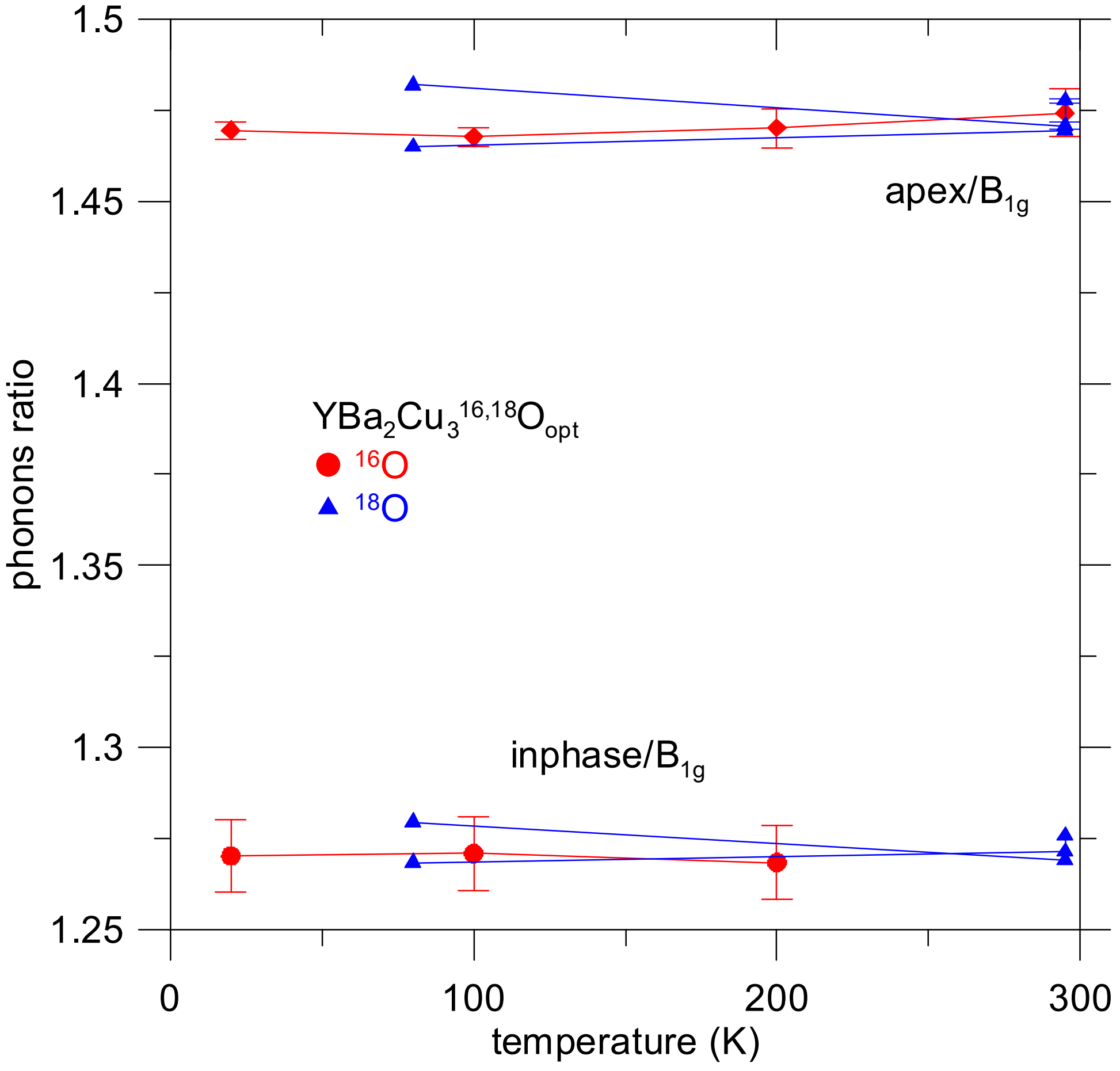

Fig.6